# Interdigitated flexible supercapacitor using activated carbon synthesized from biomass for wearable energy storage

Ankit Singh[a], Kaushik Ghosh[b], Sushil Kumar[b], Ashwini K.Agarwal[c], Manjeet Jassal[c], Pranab Goswami[a], Harsh Chaturvedi[a]*

[a]*Centre for Energy, Indian Institute of Technology, Guwahati, Assam,781039, India*
[b]*Institute of Nano Science and Technology, Mohali, Punjab,160062, India*
[c]*SMITA research lab,Indian Institute of Technology, Delhi, New Delhi, 110016, India*

**Abstract**

We have developed a flexible, interdigitated supercapacitor with high energy storage capacity for emerging wearable/flexible electronic applications. Locally obtained, low cost biomass (banana peel) was impregnated with KOH at high temperature under inert atmosphere to synthesize activated carbon with significant Brunauer–Emmett–Teller surface area of 62.03 m²/g. The supercapacitor was fabricated by screen printing interdigitated current collector of conducting silver ink on a thin flexible PET substrate and subsequent deposition of activated carbon and drop casting of gel electrolyte. Fabricated supercapacitor exhibits high capacitance of 33.18 mF/cm$^2$ at 1mV/s scan rate and 20.12 mF/cm$^2$ at discharge current of 1mA and high energy density of 5.87 μWh/cm$^2$. The developed flexible supercapacitor retains its energy storing capacity (~90%) over several cycles of mechanical bending and repetitive electronic cycling tests (~5000 cycles). Locally available biomass based activated carbon and low cost screen printing technique can be used for large scale fabrication of supercapacitor. The flexible supercapacitor demonstrates high energy storing capacity and mechanical durability through multiple bending and charging and discharging through LED. Therefore, it can be used for further developing integrated wearable and printed electronic devices.

**Introduction**

Energy storage is equally important as energy generation as generating energy at all times is neither possible nor is economically feasible. Modern day electronic devices are predominantly powered by lithium based batteries which have limitations in terms of wearability. With the rise in demand for wearable electronic devices, need for flexible/wearable energy storing device has arisen. Supercapacitor has high energy storing capacity, long life, rapid charge/discharge ability, safe from explosions and leakage and also environment friendly. [1,2] Supercapacitor can be designed to be flexible for wearable applications eliminating the necessity of carrying bulky devices. Flexible supercapacitor can also be utilized along with batteries as hybrid system for powering sensors in various areas of biomedical, environmental monitoring applications etc.[3,4,5] We have fabricated a thin flexible interdigitated supercapacitor using simple techniques which can be easily reproduced in any part of the world without the requirement of sophisticated equipment or techniques. Also, the materials used are easily available and are considerably inexpensive. [3,6,7,9]

**Experimental**

**Synthesis of activated carbon**

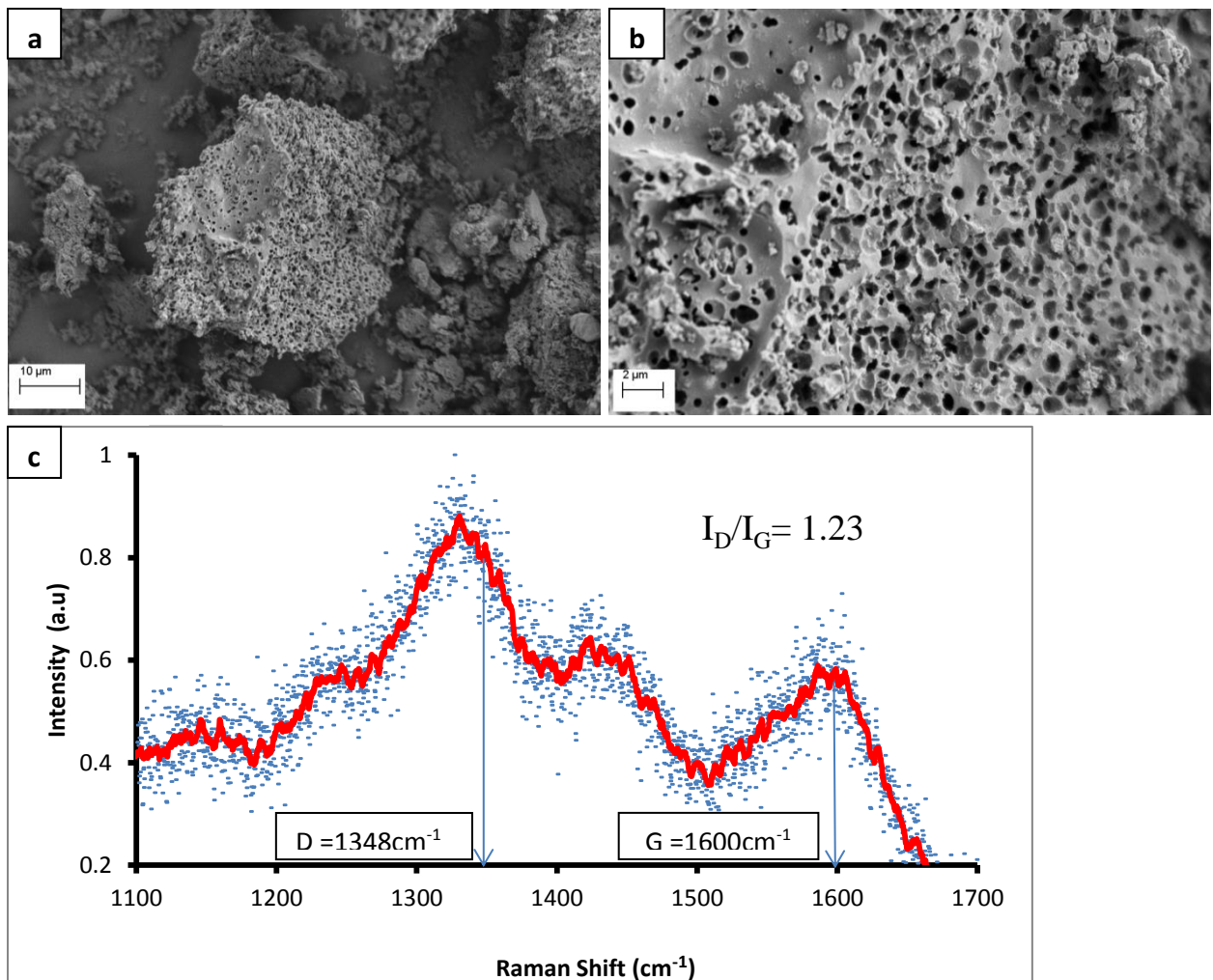

Figure 1: (a) FESEM image of synthesized activated carbon at 3kX and (b) 10kX magnification showing well porous structure, (c) Raman spectrum of activated carbon at 633nm.

Banana peel (musa acuminata) was collected from local sources. The peel was then thoroughly washed with distilled water and dried overnight in vacuum oven at 80°C.

Aqueous solution of KOH was prepared by mixing 3g of KOH flakes in 50 ml of deionized water. Small amount of dried peel (3g) was added to this aqueous KOH solution and was stirred at 200 rpm for 2 hours. The peel was removed from the solution and dried overnight in vacuum oven at 80°C. The KOH impregnated peel was heated in a muffle furnace in absence of air at 600 °C for 1 hr. The residue was washed with dilute HCL and distilled water several times until the pH reached 7 and was then dried overnight in vacuum oven at 80°C resulting in activated carbon. [7] The synthesized activated carbon was investigated using field emission scanning electron microscopy (FESEM), Brunauer–Emmett–Teller (BET) surface area analyser and Raman spectroscopy. As observed from the FESEM images (figure 1(a),(b)), the synthesized activated carbon is well porous with pore size of few hundred nanometer. The BET surface area is of 62.03m$^2$/g. The intercalation of metallic K in carbon lattices causes high microporosity of activated carbon and therefore increasing the effective surface area of the supercapacitor electrode and hence its capacitance. [7] Raman spectrum of the synthesized activated carbon was obtained by Raman spectroscopic analysis (figure 1(c)). D and G bands at 1348 cm$^{-1}$ and 1600 cm$^{-1}$ respectively obtained from the spectrum corresponds to that of activated carbon. [8] PVA/H$_3$PO$_4$ gel was prepared which acts as a separator as well as an electrolyte because it remains stable under multiple mechanical bending and has good electrochemical properties. [9]

**Preparation of gel electrolyte**

The PVA/H$_3$PO$_4$ gel electrolyte was prepared by adding 1 g of PVA into 10 ml of DI water under constant stirring for 2hrs. The whole mixture was maintained at 90°C. After a clear solution is obtained, 1ml of H$_3$PO$_4$ was added into the mixture and was stirred for another 1hr. [4,9]

**Fabrication of interdigitated supercapacitor**

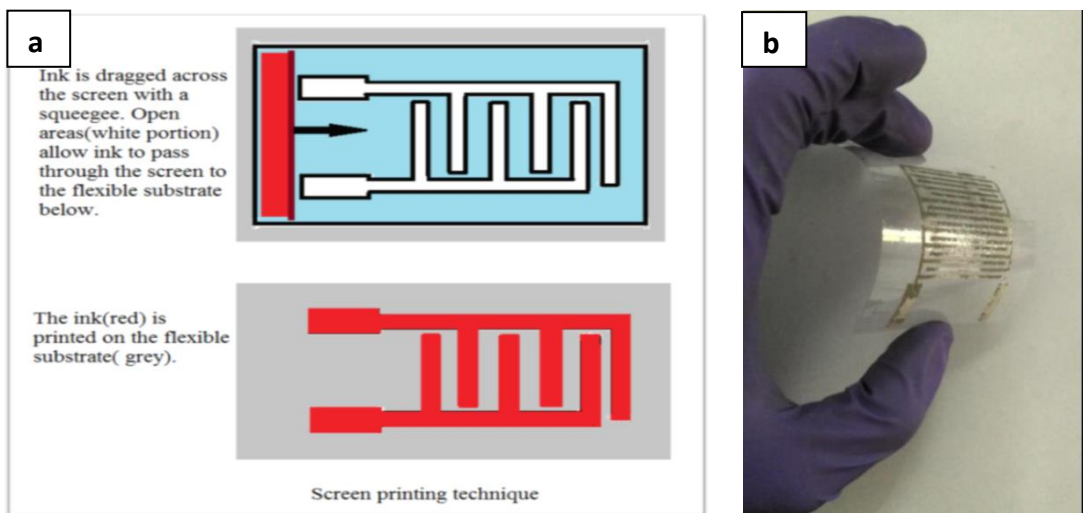

Figure 2: (a) Screen printing technique to print interdigitated current collector, (b). Silver ink based flexible interdigitated supercapacitor

Interdigitated configuration has electrodes interleaved into each other having finger like pattern (figure 2(b)). The effective capacitance increases due to electrical paralleling of individual electrodes. Acetone cleaned rectangular PET sheet (35 cm$^2$) was used as the flexible substrate. Activated carbon was crushed in mortar-pestle and was weighed and kept in equal amounts (10mg) on aluminum foil. Conductive silver ink was screen-printed on the PET substrate to form interdigitated current collector as shown in figure 2(a). Equally weighed powdered activated carbon (10mg) was evenly sprinkled on the printed current collector and was left for drying at room temperature for 5 hrs. The PVA/H$_3$PO$_4$ gel electrolyte was then drop-casted on the interdigitated electrodes and was spread using a doctor blade to get uniform thickness. The device was dried at room temperature overnight to remove excess water (figure 2(b)). [3]

**Device characterization**

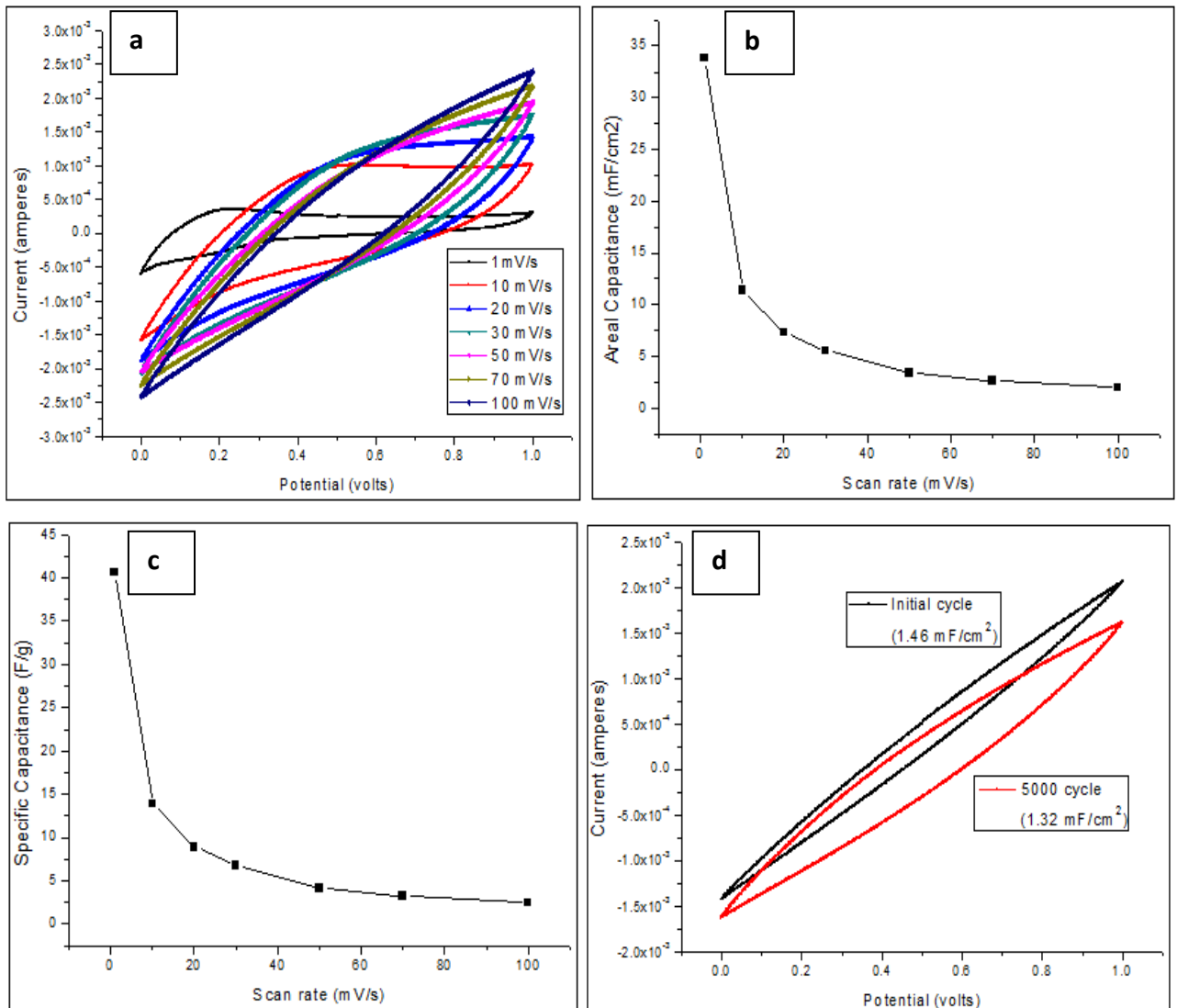

**Figure 3:**(a) CV profile of supercapacitor at different scan rate, (b) Areal capacitance plot at different scan rates, (c) Specific capacitance plot at different scan rates, (d) About 90 % of capacitance was retained after ~5000 CV cycles at 100mV/s.

Cyclic voltammetry (CV) and galvanostatic charge-discharge (CD) tests were carried out to study the electrochemical performance of the fabricated supercapacitor using Metrohm Autolab electrochemical workstation. The flexible interdigitated supercapacitor was subjected to cyclic voltammetry test at different scan rates ranging from 1mV/s to 100 mV/s in a potential window of 0V to 1V (figure 3(a)). Areal capacitance (figure 3(b)) and specific capacitance (figure 3(c)) was calculated at different scan rates using the formulae, Areal Capacitance (CV curves), $Cs = \dfrac{A}{a*v*V}$, Specific Capacitance (CV curves), $Csp = \dfrac{A}{m*v*V}$ where, A is area of CV curve, a= area of active part of device (12.25 cm$^2$), m is mass of active material (10mg), v is scan rate in mV/s and V is potential window (1V). [10] The fabricated supercapacitor retained 90% of capacitance when subjected to ~5000 CV cycles at 100mV/s to test its lifecycle (figure 3(d)).

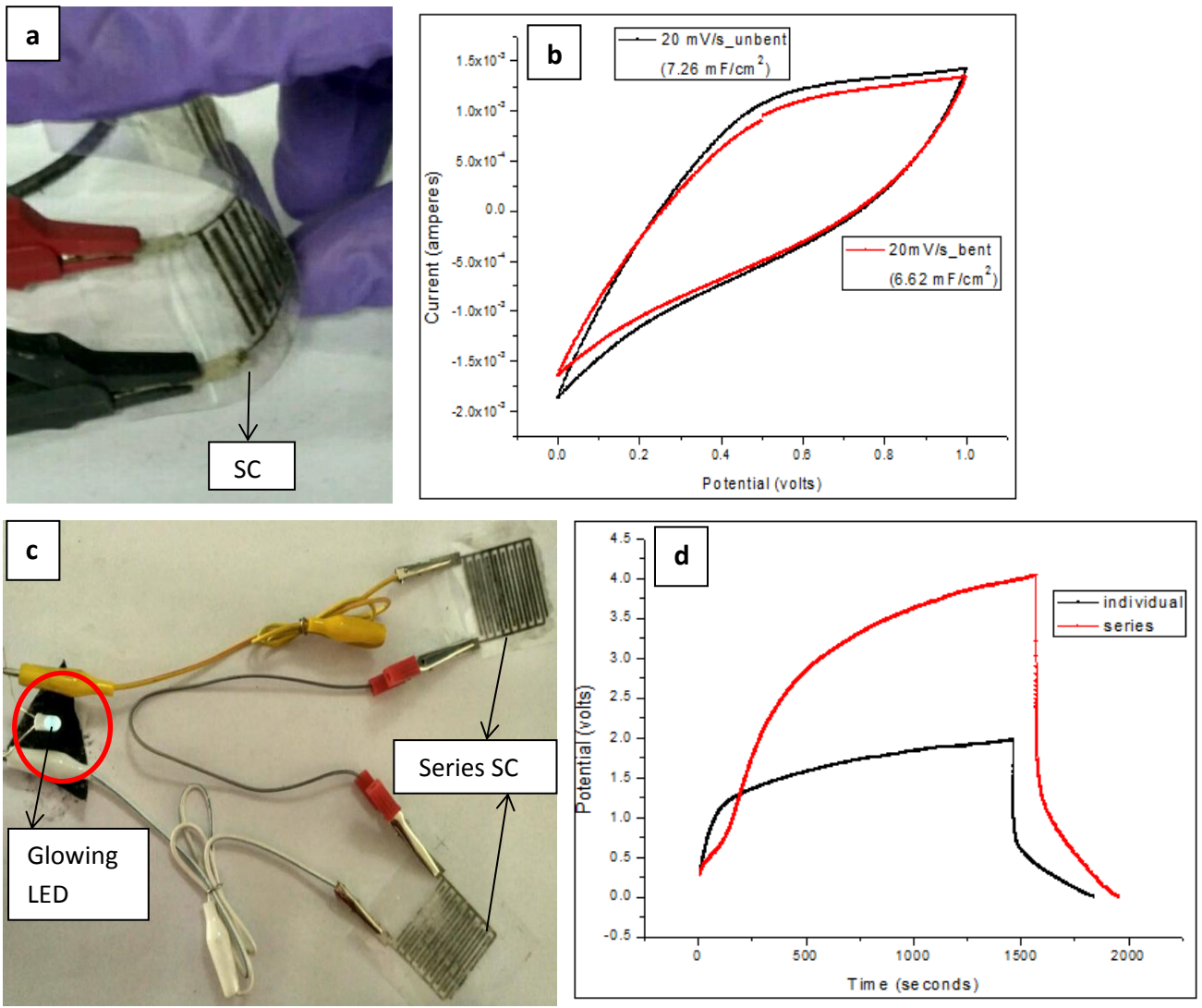

**Figure 4:** (a) Fabricated supercapacitor in bent condition, (b) Bend test at 20 mV/s caused ~ 9 % loss in capacitance, (c) Discharging series connected supercapacitor through LED, (d) CD test at 1mA of series and individual supercapacitor

The fabricated supercapacitor (SC) was subjected to CV test at 20 mV/s under unbent and bent condition (Figure 4(a),(b)). The supercapacitor retained about 91% of capacitance during bending condition showing it can be used for flexible applications. Series connected supercapacitor were charged at 5V using DC power supply for 10 minutes and discharged through a LED. The LED glowed for 2 minutes after which the voltage reduced below the threshold value of 2.3V for white LED (figure 4(c)). CD tests were performed at 1mA on individual and series connected supercapacitor (figure 4(d)). From the CD curve, individual supercapacitor attained a saturation voltage of 1.97 V. Areal capacitance (CD profile) of individual supercapacitor was calculated using, $Cs = \frac{I * td}{a * V'}$ where, I is the constant discharge current, td is discharge time, a= area of active part of device (12.25 cm$^2$), V' is effective discharge voltage (1.45 V, excluding IR drop), which was equal to 20.12 mF/cm$^2$. Also the energy and power density of individual supercapacitor was calculated using Energy density, $Es = \frac{0.5 * Cs * V'^2}{3600}$ and Power density, $Ps = \frac{Es * 3600}{td}$ which were equal to 5.87 µWh/cm$^2$ and 59.18 µW/cm$^2$ respectively. Equivalent series resistance (ESR) was calculated from CD curve using $ESR = \frac{V''-V'}{I}$ where, V" is the saturation voltage (1.97 V), which equals to 520 Ω. High value of ESR can be due to high mass loading (10mg) of active electrode material. [10]

**Conclusion**

This work demonstrates a durable thin flexible interdigitated supercapacitor fabricated using simple techniques/methods and easily available inexpensive materials which reduces the overall cost of the device considerably. The fabricated supercapacitor has demonstrated high capacitance of 33.18 mF/cm$^2$ at 1mV/s scan rate and 20.12 mF/cm$^2$ at discharge current of 1mA (~ 82µA/cm$^2$) and high energy density of 5.87µWh/cm$^2$. The flexible supercapacitor showed a satisfactory performance retaining ~ 90% of energy storing capacity after bending and multiple electronic cycling tests (~5000 cycles). Despite the good performance, silver ink based current collector gets degraded over time as it reacts with the acid present in the gel electrolyte. So, conductive ink based on carbon material can be used to print electrodes which would tolerate the acid for longer period and will further reduce the cost of the device. Activated carbon could be used along with conductive carbon material and nanosized transition metal oxides as electrode material to further enhance the performance of the supercapacitor. Hence, the developed thin flexible interdigitated supercapacitor with high energy storing capacity can be a better alternative as a wearable energy source. Also, the fabricated supercapacitor can be utilized for further development of emerging flexible/wearable electronic devices.

**Acknowledgements**

We would like to thank Centre for Energy, Central Instruments Facility of Indian Institute Of Technology, Guwahati. We would also like to thank G Labs Pvt ltd, Kolkata for their support.

**References**

[1] US 2800616, Becker, H.I., "Low voltage electrolytic capacitor", issued 1957-07-23

[2] B.E.Conway. Transition from Supercapacitor to Battery Behaviour in Electrochemical Energy storage. *J.Electrochem.Soc*.138(6): 1539-1548.

[3] R. Kumar, R. Savu, E. Joanni et al. Fabrication of interdigitated micro-supercapacitor devices by direct laser writing onto ultra-thin, flexible and free-standing graphite oxide films. *RSC Adv*, 6, (2016), 84769.

[4] Z. Niu, L. Zhang, L. Liu et al. All-Solid-State Flexible Ultrathin Micro-Supercapacitors Based on Graphene. *Adv. Mater*, 25, (2013), 4035–4042.

[5] X. Wang, K. Jiang, G. Shen. Flexible fiber energy storage and integrated devices: recent progress and perspectives. *Materials Today*, 18, 5, (2015).

[6] Lewandowski, A. Zajder, M. Frąckowiak et al. Supercapacitor based on activated carbon and polyethylene oxide–KOH–$H_2O$ polymer electrolyte. *Electrochimica Acta*, 46, 18, (2001), 2777-2780.

[7] Wang, J. Kaskel, Stefan. KOH activation of carbon-based materials for energy storage. *J. Mater. Chem.*, 22, (2012), 23710-23725.

[8] N. Shimodaira *and* A. Masui. Raman spectroscopic investigations of activated carbon materials. *Journal of Applied Physics*, 92, 902, (2002).

[9] Chen, Q. Li, X. Zang et al. Effect of different gel electrolytes on graphene-based solid-state supercapacitors. *RSC Adv*, 4, 68, (2014), 36253-36256.

[10] B.D. Boruah, A. Maji, A. Misra. . Flexible Array of Microsupercapacitor for Additive Energy Storage Performance Over a Large Area. *ACS Appl. Mater. Interfaces*, 10, (2018), 15864−15872.